\documentclass[%
 reprint,
 amsmath,amssymb,
 aps,
]{revtex4-1}

\usepackage{graphicx}
\usepackage{dcolumn}
\usepackage{bm}
\usepackage{braket}
\usepackage{amsmath}
\usepackage{epstopdf}
\usepackage{siunitx}
\usepackage{amsfonts}
\usepackage{float} 

\begin{document}

\preprint{APS/123-QED}

\title{Large Perturbation Method}

\author{G. Mikaberidze}%
 \email{mikaberidze@ucdavis.edu}
\affiliation{
Department of Physics, University of California, Davis, CA, 95616, USA
}
\affiliation{
School of Physics, Free university of Tbilisi, Tbilisi 0131, Georgia
}

\date{\today}

\begin{abstract}
This paper describes a new numerical method for solving eigenstate problems, such as time-independent Schrodinger equation. The idea is to use the first order perturbation theory to rewrite the eigenvalue problem as a system of first order differential equations and then solve them using numerical techniques. The method allows to introduce perturbation terms of any order of magnitude. The algorithm is in some cases faster than conventional variational method and offers a new insight into perturbation theory. It is also easy to understand and implement.
\end{abstract}

\pacs{Valid PACS appear here}
\maketitle


\section{\label{sec:level1}Introduction}

Standard perturbation theory does not work very well for large interaction terms, because usually it produces divergent series \cite{BenderWu}. The standard approach is to use various summation techniques to extract answers from these divergent series \cite{bookInAsymptotic}, although it has not been shown yet, that this is always possible \cite{ptacs}. The method described in this paper is interesting, as it presents a different way of approaching the problem.

We will first describe the method and derive it from the standard perturbation theory. We will see, that in some cases, this method is faster then conventional numerical method, variational principle. Then we will calculate the efficiency of the algorithm and reformulate it in a new and faster way. The results are examined with two examples: 1. Anharmonic oscillator with large quartic term. We will compare our results to the results calculated through other techniques; 2. Using infinite well as unperturbed problem, we will numerically calculate eigenvalues and eigenfunctions for Poschl-Teller potential. As both, unperturbed and perturbed problems are analytically solvable, this will allow us to compare the numerical and analytical solutions.

The method described in this paper is a numerical generalisation of perturbation theory. The perturbation theory can be applied to the problems, which differ from some exactly solvable problem by a "small" term. The advantage of this method is that it allows the difference to be any finite term.

The main idea is that we can divide the perturbation term by a large number $N$ and then perturb the initial solution step by step $N$ times using the obtained small perturbation. This is actually a simplification of the real idea. To be more specific, the above explanation compares to the Euler method for solving ODEs, whereas the real algorithm allows for using the Runge-Kutta-Fehlberg or any other convenient method.

\section{derivation and description of the method}
We can write the ordinary perturbation theory in the following form 

$$\hat{H}^{(0)}\ket{n^{(0)}}=E_n^{(0)}\ket{n^{(0)}}$$
$$\hat{H}^{(1)}= \hat{H}^{(0)}+\epsilon \hat{H}$$
$$E_n^{(1)}=E_n^{(0)}+\epsilon\dot{E}_n^{(0)}+O(\epsilon^2)$$
$$\ket{n^{(1)}}=\ket{n^{(0)}}+\epsilon\dot{\ket{n^{(0)}}}+O(\epsilon^2)$$
$$ \dot{E}_n^{(0)}=\braket{n^{(0)}|\hat{H}|n^{(0)}}$$
$$\dot{\ket{n^{(0)}}}=\sum_{i\ne n}\ket{i^{(0)}}\dfrac{\braket{i^{(0)}|\hat{H}|n^{(0)}}}{E_n^{(0)}-E_i^{(0)}}$$

Where $\hat{H}^{(0)}$ represents the Hamiltonian of the solvable system; $E_n^{(0)}$ and $\ket{n^{(0)}}$ are $n$th eigenvalue and eigenfunction of the given Hamiltonian; $\hat{H}^{(1)}$ is the perturbed Hamiltonian and $\epsilon \hat{H}$ is the perturbation term; naturally $E_n^{(1)}$ and $\ket{n^{(1)}}$ are $n$th eigenvalue and eigenfunction of the perturbed Hamiltonian.

We have obtained all the information about the eigenstates of the perturbed Hammiltonian, now we can apply the same method once more and perturb the Hamiltonian a bit more. In general we can use the method as many times as we want. If we take $\epsilon=1/N$ where $N$ is a large integer and use the perturbation theory $N$ times, finally we will arrive at the solution for $\hat{H}^{(N)}= \hat{H}^{(0)}+\hat{H}$. The recurrent form of the equations is as follows

$$\hat{H}^{(k)}= \hat{H}^{(0)}+k\epsilon \hat{H}$$
\begin{equation}
    E_n^{(k+1)}=E_n^{(k)}+\epsilon\dot{E}_n^{(k)}+O(\epsilon^2)
    \label{EDiscrete}
\end{equation}
\begin{equation}
    \ket{n^{(k+1)}}=\ket{n^{(k)}}+\epsilon\dot{\ket{n^{(k)}}}+O(\epsilon^2)
    \label{psiDiscrete}
\end{equation}
\begin{equation}
    \dot{E}_n^{(k)}=\braket{n^{(k)}|\hat{H}|n^{(k)}}
    \label{EDiscreteDer}
\end{equation}
\begin{equation}
    \dot{\ket{n^{(k)}}}=\sum_{i\ne n}\ket{i^{(k)}}\dfrac{\braket{i^{(k)}|\hat{H}|n^{(k)}}}{E_n^{(k)}-E_i^{(k)}}
    \label{psiDiscreteDer}
\end{equation}

Where $\hat{H}^{(k)}$ is the Hamiltonian after $k$ iterations and $E_n^{(k)}$ and $\ket{n^{(k)}}$ are its eigenvalues and eigenfunctions.

Let us now expand the eigenfunctions $\ket{n^{(k)}}$ in basis of eigenfunctions of the initial Hamiltonian

\begin{equation}
    \ket{n^{(k)}}=\sum_j Q_n^{k j}\ket{j^{(0)}}\equiv\sum_j Q_n^{k j}\ket{j}
    \label{expansionDiscrete}
\end{equation}

In case of degeneracy the basis should be chosen in such a way, that the perturbed Hamiltonian would be diagonal in this degenerate subspace. In other words, the degeneracy is resolved using standard approach. It is obvious that $Q_n^{0 j}=\delta_n^j$. Now we will rewrite the equations (\ref{psiDiscrete})-(\ref{psiDiscreteDer}) using equation (\ref{expansionDiscrete}). The matrix element of perturbation term would be

\begin{align}
    \label{matrixElement}
    \begin{split}
        \braket{i^{(k)}|\hat{H}|n^{(k)}}=\sum_jQ_i^{kj*}\sum_pQ_n^{k p}\braket{j|\hat{H}|p}\equiv \\
        \equiv\sum_{j,p}Q_i^{kj*}Q_n^{k p}H_{jp}
    \end{split}
\end{align}

Using equations (\ref{EDiscrete}), (\ref{EDiscreteDer}) and (\ref{matrixElement}), we obtain

\begin{equation}
    \label{EnDiscreteFinal}
    E_n^{(k+1)}=E_n^{(k)}+\epsilon\sum_{j,p}Q_n^{kj*}Q_n^{k p}H_{jp}
\end{equation}

Substituting (\ref{psiDiscreteDer}), (\ref{expansionDiscrete}) and (\ref{matrixElement}) into (\ref{psiDiscrete}) and multiplying the result by some $\bra{q}$ we obtain

\begin{equation}
    \label{QnDiscreteFinal}
    Q_n^{(k+1) q}=Q_n^{k q}+\epsilon\sum_{i\ne n,j,p}\dfrac{Q_i^{k q}Q_i^{k j *}Q_n^{k p}}{E_n^{(k)}-E_i^{(k)}} H_{jp}
\end{equation}

Now instead of a discrete $k$ we introduce a continuous variable $\sigma$ which takes values from 0 to 1. Each value of $\sigma$ corresponds to a different Hamiltonian $\hat{H}^{(1)}(\sigma)= \hat{H}^{(0)}+\sigma \hat{H}$. The small variable $\epsilon$ is changed to $d\sigma$. 
It is easy to show that equations (\ref{EnDiscreteFinal}) and (\ref{QnDiscreteFinal}) can be rewritten as a set of first order differential equations

\begin{equation}
    \label{difE}
    \dfrac{dE_n(\sigma)}{d\sigma}=\sum_{j,p}Q_n^{j *}(\sigma)Q_n^p(\sigma)H_{jp}
\end{equation}

\begin{equation}
    \label{difQ}
     \dfrac{dQ_n^q(\sigma)}{d\sigma}=\sum_{i\ne n,j,p}\dfrac{Q_i^q(\sigma)Q_i^{j *}(\sigma)Q_n^p(\sigma)}{E_n(\sigma)-E_i(\sigma)} H_{jp}
\end{equation}

So it is reduced to an initial value problem, with initial conditions $Q_n^j(0)=\delta_n^j$, $E_n(0)=E_n^{(0)}$ and $H_{jp}=\braket{j|\hat{H}|p}$. 

The sums here can be thought of as generalized sums (including integration), if the spectrum of unperturbed Hamiltonian has continuous part too.

The equations (\ref{difE}) and (\ref{difQ}) contain infinite sums. In order to calculate the approximate values for them, the immediate approach is to only consider first $N$ terms and neglect the others. If we take big enough $N$, the approximation will differ from the real infinite sum by a small term. On the other hand, in contrast to variational principle, to calculate perturbation for $n$th eigenstate, with large number $n$, one does not need to consider all the eigenstates smaller then $n$. It is enough to consider $N$ neighbouring eigenstates and neglect others, as the contribution of $i$th eigenstate ($i\ne n$) on the perturbation of the $n$th eigenfunction is inversely proportional to $E_n-E_i$. For large enough $n$, this method will be faster than conventional variational method.

A possible way to control the error of calculations is to use a suitable acceleration technique on the sums. One should calculate the results using standard sums and afterwards calculate more precise results using accelerated sums. The absolute value of difference between these two should be of the same order as the error of the calculations involving the standard sums.

\section{algorithm efficiency}

In this section we will estimate and improve the efficiency of the algorithm. 

To proceed with this method, first of all, we will have to prepare the $H_{jp}$ matrix elements. If we consider $N$ eigenstates, this calculations will take $O(N^2)$ operations. Although the rest of the algorithm has higher asymptotic time in terms of $N$, these operations involve integration, and for some practical problems might be the slowest part of the calculations.

After this, in the process of solving the ODEs (\ref{difE}) and (\ref{difQ}), we will have to evaluate the right hand sides of the equations. Each evaluation of the equation (\ref{difE}) will take $O(N^2)$ operations, as it involves double sum over $N$ terms. We will have to calculate derivatives for $N$ eigenvalues, thus the energies will take $O(N^3)$ operations altogether.

Now let us estimate the number of operations needed to evaluate (\ref{difQ}). There are $N$ eigenstates, each eigenfunction has $N$ components. Thus there are $N^2$ $Q$ coefficients. If you implement the formulae right away, each calculation will require $O(N^3)$ operations, as it involves triple sum, and altogether, $Q$s will use $O(N^5)$ operations, which would be slow.

To improve the algorithm, we need to rewrite the equation (\ref{difQ}) in the following way

$$\dfrac{dQ_n^q}{d\sigma}=\sum_{i\ne n,j,p}\dfrac{Q_i^qQ_i^{j *}Q_n^p}{E_n-E_i} H_{jp}=\sum_{i\ne n}Q_i^qT_{in}$$

$$T_{in}\equiv\sum_{j,p}\dfrac{Q_i^{j *}Q_n^p}{E_n-E_i} H_{jp}=\sum_{j}\dfrac{Q_i^{j *}}{E_n-E_i} S_{nj}$$

$$S_{nj}\equiv\sum_pQ_n^pH_{jp}$$

Each element of $S_{nj}$ requires $O(N)$ operations, thus for the whole $S_{nj}$, it will be $O(N^3)$. Each element of $T_{in}$ will take $O(N)$ operations and the whole $T_{in}$ will need $O(N^3)$. After this, we have $T_{in}$ and calculation of each $\dfrac{dQ_n^q}{d\sigma}$ will only take $O(N)$ operations, altogether $O(N^3)$ again. Therefore, the whole process will take $4 O(N^3)=O(N^3)$ operations, which is a great improvement. One should note, that $N$ here is not the largest number of eigenstate, but rather the total number of eigenstates considered.

To accelerate the algorithm even more, one could make use of parallel programming: calculate each term of the sum in parallel, then summ by pairs in parallel again and so on recursively. In this case the number of required consecutive operations would be of order $O(\log{N})$.

\section{examples}

\subsection{Anharmonic oscillator}

In this example we are going to introduce a quartic term in the Hamiltonian for a harmonic oscillator. This has become a standard test case for similar purposes.

$$\hat{H}^{(1)}(\sigma) = \hat{H}^{(0)}+\sigma \hat{H}$$
$$\hat{H}^{(0)} = -\dfrac{\partial^2}{\partial x^2}+x^2$$
$$\hat{H} = x^4$$

\begin{figure}[h]
    \centering 
    \includegraphics[scale=1]{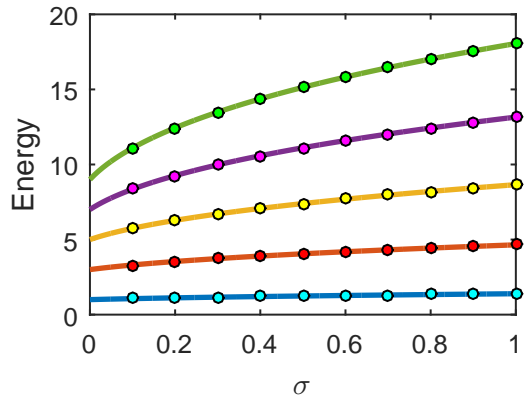}
    \caption
    {\label{move} 
        Energy vs $\sigma$ for first five eigenstates.
        The lines are results obtained with this method. The points correspond to the values calculated by J. E. Drummond \cite{quartic}.
    }
    \label{anharmonicFig}
\end{figure}

Knowing the eigenfunctions of the unperturbed problem, we calculate the matrix elements $H_{jp}$ and then solve equations (\ref{difE}) and (\ref{difQ}) numerically using Runge-Kutta-Fehlberg method, taking into account only first 30 eigenstates. Figure \ref{anharmonicFig} represents the comparisonof our results to the values calculated by J. E. Drummond using technique to sum the divergent series \cite{quartic}. Our method reproduced every digit in the table 3 given in \cite{quartic}.

\subsection{From infinite well to Poschl-Teller potential}

In this example we will use our method to perturb infinite well solutions with Poschl-Teller potential and obtain the eigenstates for it.

This example has been chosen because there exist analytical solutions for both, unperturbed and perturbed systems and because the perturbation term is obviously not small.

$$\hat{H}^{(1)}(\sigma)= \hat{H}^{(0)}+\sigma \hat{H}$$

$$\hat{H}^{(0)} = -\dfrac{\partial^2}{\partial x^2}+V_1(x)$$
$$
    V_1(x)=\begin{cases}
        0, & \text{if $|x|<\pi/2$}.\\
        \infty, & \text{otherwise}.
    \end{cases}
$$
$$
    \hat{H} =\begin{cases}
        \dfrac{100}{\cos^2{x}}, & \text{if $|x|<\pi/2$}.\\
        \infty, & \text{otherwise}.
    \end{cases}
$$

\begin{figure}[H]
    \centering 
    \includegraphics[scale=0.9]{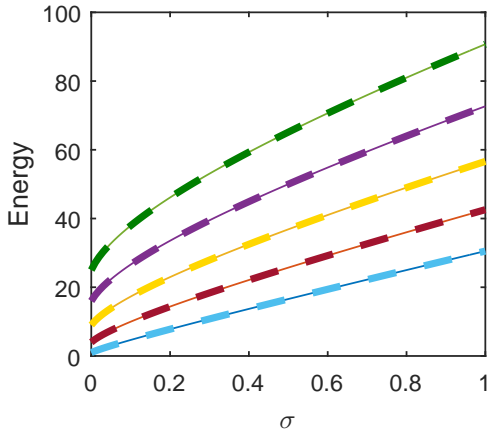}
    \caption
    {\label{move} 
        Energy vs $\sigma$ for first five eigenstates.
        Thin lines are the analytical results, whereas thick dash represents the numerical results. 
    }
    \label{tellerFig}
\end{figure}

For these expressions we numerically solve equations (\ref{difE}) and (\ref{difQ}) much like the previous example. In these calculations 30 eigenstates were considered. In figure \ref{tellerFig} you can see the dependence of first five eigenvalues on the $\sigma$ parameter. The errors are larger for the eigenstates close to upper limit.

\begin{figure}[H]
    \centering 
    \includegraphics[scale=0.9]{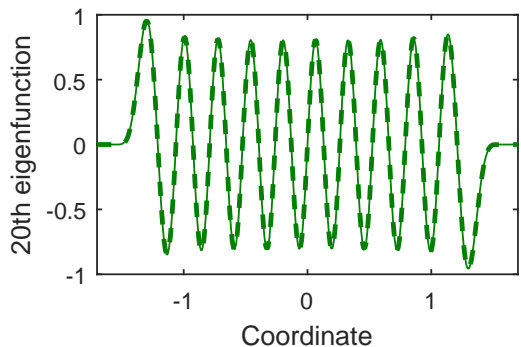}
    \caption
    {\label{move} 
        Numerical and analytical solutions for the $20$th eigenfunction for fully perturbed problem. Thin  line is the analytical curve, whereas thick dash represents the numerical result (the number of eigenvalue was chosen randomly).
    }
    \label{tellerEigenFig}
\end{figure}

Having equation (\ref{expansionDiscrete}) and knowing values $Q_n^q(\sigma)$, we can reconstruct the eigenfunctions for any given value of $\sigma$ (see figure \ref{tellerEigenFig}).

\section{conclusions}


Perturbation theory stays an important tool for approximately solving problems in quantum mechanics \cite{pertBook}. It plays a very important role for example in in quantum field theory \cite{feynman}. 

We have modified the eigenvalue problem to become an initial value problem, which can be solved approximately using any standard techniques for solving a system of ordinary differential equations. This method, in some cases has a better efficiency than the conventional variational method.

The method has been tested for anharmonic oscillator and the results matched the ones obtained in publication \cite{quartic} through different technique. Another test was to start with one analytically solvable system (infinite well problem), add to the Hamiltonian a term to turn it into another analytically solvable system (particle in Poschl-Teller potential). The results matched in this case too.

Therefore we can say, that we have derived a method to approach eigenvalue problems with finite perturbation terms of any order.

\section{acknowledgements}
The author wants to thank Professor George Jorjadze for many valuable and interesting discussions and also the University of Zielona Gora for the hospitality provided during the work on this paper.

\end{document}